\documentclass[preprint,12pt]{article}
\usepackage{amssymb}
\usepackage{graphicx}
\usepackage{multirow}
\usepackage{caption}
\usepackage{subfig}
\begin{document}
\title
{General relativity and relativistic astrophysics}
\author{Banibrata Mukhopadhyay\\ \\
Department of Physics, Indian Institute of
  Science, Bangalore 560012, India\\
bm@physics.iisc.ernet.in}
\maketitle
\begin{center}
Article published for a special section in Current Science dedicated to 100 years
of general relativity
\end{center}

\vskip1.0cm
\begin{abstract}

Einstein established the theory of general relativity and the corresponding 
field equation in 1915 and its vacuum solutions were 
obtained by Schwarzschild and Kerr for, respectively, static and rotating black holes, in 
1916 and 1963, respectively. They are, however, still playing an indispensable role, even after 100 years 
of their original discovery, to explain high energy astrophysical phenomena. Application of 
the solutions of Einstein's equation to resolve astrophysical phenomena has formed an important
branch, namely relativistic astrophysics. I devote this article to enlightening some of the 
current  astrophysical problems based on general relativity. However, there seem to be some 
issues with regard to explaining certain astrophysical phenomena based on Einstein's theory alone. I show that 
Einstein's theory and its modified form, both are necessary to explain modern astrophysical processes,
in particular, those related to compact objects.

\end{abstract} 


\newpage

\section{Introduction}

Within a few months of the celebrated discovery of Einstein's field equation \cite{ein},
Schwarzschild obtained its vacuum solution in spherical symmetry \cite{sch}. However, it
took almost another half a century before Kerr obtained its vacuum solution for an
axisymmetric spacetime \cite{kerr}, which was a very complicated job at that time. The former
solution is very useful to understand the spacetime properties around a static black hole,
called Schwarzschild black hole. The latter solution corresponds to the spacetime properties around a 
rotating black hole, called Kerr black hole, in particular after its generalization by Boyer and Lindquist \cite{bl} 
to its maximal analytic extension. Both the solutions have enormous applications to relativistic
astrophysics; however, as black holes in general possess spin, the Kerr solution
is much more important. In the Boyer-Lindquist coordinates, the outer radius of 
a black hole is defined as $r_+=GM/c^2\left(1+\sqrt{1-a^2}\right)$,
when $M$ is the mass of the black hole, $c$ the speed of light, $G$ the Newton's gravitational constant and
`$a$' the spin parameter (angular momentum per unit mass) of
the black hole. Hence, for $|a|> 1$, the collapsed
object will form a naked singularity without an event horizon, rather than a black hole.
In addition, $a=0$ corresponds to the Schwarzschild black hole. 
Hence, predicting `$a$' of
black holes from observed data would serve as a natural proof for the existence of the Kerr metric in the universe.

In the presence of matter (i.e. nonvanishing energy-momentun tensor of the source field
$T_{\mu\nu}$), there are a 
variety of solutions of Einstein's equations (e.g. \cite{hr,cook,xns}), depending upon the equation of state (EoS). In order
to understand the properties of neutron stars, and also white dwarfs, these solutions serve as very important tools.
In this context, a very important class of objects is binary pulsars, which are one of the few objects 
that help to test Einstein's general relativity (GR). Such binary systems have a pulsating star along with a companion, 
often a white dwarf or a neutron star.
PSR~B1913+16 was the first binary pulsar discovered by J. Taylor and R. Hulse
which led to them wining the Nobel Prize in Physics in 1993 \cite{hulse}. It has been found that its
pulsating rate varies regularly due to the Doppler effect, when it is orbiting another star very closely 
at a high velocity. PSR~B1913+16 also allowed determining accurately the masses of neutron stars, using relativistic 
timing effects. When the two components of the binary system are coming closer, the gravitational field appears to be 
stronger and, hence,  creating time delays which furthermore in turn increase pulse period.
Binary pulsars, as of now, are perhaps the only tools based on which gravitational waves are being evident. 
According to GR, two neutron stars in a binary system would emit gravitational waves while 
orbiting a common center of mass and, hence, carrying away orbital energy. As a result, the two stars  come closer 
together, shortening their orbital period, which we observe.

Although the validity of the solutions of Einstein's equation, i.e. GR,
has been well tested, particularly in the weak field regime --- such as
through laboratory experiments and solar system tests --- question remains, whether GR is the ultimate theory 
of gravitation or it requires modification in the strong gravity regime. 
Indeed, scientists have been trying to resolve the astrophysical problems related to the strong field regimes, 
like expanding universe, massive neutron stars, by introducing modified theories of GR
(e.g. \cite{staro,capo,eksi}). Recently, there are observational evidences for
massive neutron star binary pulsars PSR~J1614-2230 \cite{nat} and  PSR~J0348+0432 \cite{sc}
with masses $1.97M_\odot$ and $2.01M_\odot$ respectively, where $M_\odot$ is solar mass. 
Similarly, there is a lot of interest in exploring massive white dwarfs (see \S4 for details).
The possibility of very massive neutron stars has been examined \cite{weissenborn} in the presence of
hyperons and the conditions to obtain the same. Note that the likely presence of $\Lambda$-baryons in dense hadronic 
matter tends to soften EoS such that the above mentioned massive neutron stars are difficult 
to explain, known as `hyperon problem'. Based on the quark-meson coupling model, 
it has been shown \cite{whittenbuary} that the maximum mass of neutron stars could be $\approx 2M_\odot$,
when nuclear matter is in $\beta$-equilibrium and hyperons must appear. Apart from 
the EoS based exploration, neutron stars with mass $\gtrsim 2M_\odot$ have been shown to be possible
by exploring effects of magnetic fields, with central field $\sim 10^{16}$G \cite{pili}, and modification 
to GR \cite{capo,eksi,cheon}.

Black holes are not visible and neutron stars too are hardly visible, unless the latter possess stronger 
magnetic fields. Hence, in order to understand their properties, light coming out off the matter infalling 
towards them (as well as influenced by them), called accretion, plays a very important role. Study of accretion
around compact objects is a vast part of relativistic astrophysics. While a simple spherical accretion
model in the Newtonian framework was introduced by Bondi in the 50s \cite{bondi}, later its general relativistic version was
worked out by Michael \cite{michael} in the Schwarzschild spacetime, which was perhaps the first venture 
into accretion physics in GR. However, generically, accretion flows possess angular momentum, as inferred
from observed data, forming accretion disks around compact objects. Such a (Keplerian) disk model in the general relativistic framework was 
formulated by Novikov and Thorne \cite{nt73} (whose Newtonian version \cite{ss73} is highly popular as well).
Later on, in order to satisfactorily explain observed hard X-rays, the geometrically thick 
(and sub-Keplerian) disk model was initiated, in the Newtonian (e.g. \cite{sel76,ny94}), pseudo-Newtonian
(e.g. \cite{pw80,m02}), as well as
general relativistic (e.g. \cite{liang80,c90,gp98,belo07}) frameworks. All of them explicitly reveal the
importance of GR in accretion flows.

Furthermore, observed jets from black hole sources have been demonstrated to be governed by
general relativistic effects in accretion-outflow/jet systems, based on general relativistic magnetohydrodynamic (GRMHD)
simulations, with and without the effects of radiation (e.g. \cite{sasha11,jon12,jon14}). It
has been demonstrated therein that the spin of black holes plays a crucial role to control the underlying
processes. It is also known that accretion flows (directly or indirectly) are intertwined with several other observed 
relativistic features in modern astrophysics, e.g. quasi-periodic oscillation (QPO) in compact sources,
gamma-ray bursts (combined disk-jet systems), supernovae etc. In recent years,
many observations reveal that several gamma-ray bursts (which are the extremely energetic explosions that 
have been observed in distant galaxies) occur in coincidence
with core-collapse supernovae, which are related to the formation of black holes and neutron stars.

American federal institutions such as NASA, European agencies such as
ESO, Japanese institutions etc. have been devoted to conduct numerous
satellite experiments (such as HST, Chandra, XMM-Newton, Swift, Fermi, Astro-H, Suzaku etc.) which
regularly receive data from galactic and extragalactic (compact) sources, producing all the above mentioned
features. Similarly, Indian satellite Astrosat is gathering data from black hole, white dwarf and neutron star sources.
All these missions help in understanding relativistic astrophysical sources, their evolution and up-to-date
status. They furthermore help to verify theoretical concepts of GR.

In the present article, I plan to touch upon some of the specific 
issues in relativistic astrophysics, the ones which are very {\it hot-topics} at present and I am working on them, 
in detail. However,
before I go into their detailed discussions, in the next section, let me recall some of their basic building blocks.

\section{Some basic formulation}

Let me start with the 4-dimensional action as \cite{livrel}
\begin{equation}
S = \int \left[\frac{1}{16\pi} f(R) + {\cal L}_M \right] \sqrt{-g}~ d^4 x , 
\label{action}
\end{equation}
where $g$ is the determinant of the spacetime metric $g_{\mu\nu}$, 
${\cal L}_M$ the Lagrangian density of the matter field, 
$R$ the scalar curvature defined as $R=g^{\mu\nu}R_{\mu\nu}$, 
where $R_{\mu\nu}$ is the Ricci tensor and $f$ is an 
arbitrary function of $R$; in GR, $f(R)=R$. 
Now, on extremizing the above action for GR, one obtains Einstein's field equation as
\begin{equation}
G_{\mu\nu}= R_{\mu\nu}-\frac{g_{\mu\nu }R}{2}=\frac{8\pi G T_{\mu\nu}}{c^4},
\label{fld}
\end{equation}
where $G_{\mu\nu}$ is called the Einstein's field tensor.

For black holes,  $T_{\mu\nu}=0$ and, hence, the spacetime metric for the vacuum solution
of a charged, rotating black hole (Kerr-Newman black hole) with $G=c=1$ in the Boyer-Lindquist coordinates is
\begin{eqnarray}
\nonumber
ds^2 = \frac{(\Delta-a^2\sin^2\theta)}{\tilde{\rho}^2}dt^2-\frac{\tilde{\rho}^2}{\Delta}dr^2-\tilde{\rho}^2 d\theta^2
-\left[(r^2+a^2)^2-a^2\Delta\sin^2\theta\right]\frac{\sin^2\theta}{\tilde{\rho}^2}d\phi^2\\
-\left(\Delta-r^2-a^2\right)\frac{2a\sin^2\theta}{\tilde{\rho}^2}dt d\phi,
\label{kerrm}
\end{eqnarray}
where $\Delta=r^2-2Mr+a^2+Q^2$, $\tilde{\rho}^2=r^2+a^2\cos^2\theta$, 
$Q$ is the charge per unit mass of the black hole. 
For $Q=0$, the metric equation (\ref{kerrm}) reduces to the Kerr metric (rotating uncharged black hole), for $a=Q=0$ it reduces
to the Schwarzschild metric (non-rotating uncharged black hole) and for $a=0$ it reduces to the Reissner-Nordstr\"om 
metric (charged non-rotating black hole).

For a neutron star (as well as white dwarf which, however, generally could be explained mostly
by the Newtonian theory, except the cases described in \S4), $T_{\mu\nu}\neq 0$ and 
the solution of equation (\ref{fld}) depends on EoS and in general there 
is no analytic solution \cite{cook,xns} (see, however, \cite{hr}). Therefore, for most commonly observed stationary,
axisymmetric (rotating) neutron stars and white dwarfs, 
$g_{tt}$, $g_{rr}$, $g_{\theta\theta}$, $g_{\phi\phi}$ and $g_{t\phi}$
could be obtained as numerical functions of $r$ and $\theta$ (e.g. \cite{xns}), and $M(r)$ therein would be interpreted
as the mass enclosed in the star upto the radial distance $r$ from the center.

In order to obtain the solutions of accreting matter around a black hole and stellar structure for 
a neutron star and a white dwarf, one has to solve the stress-energy tensor equation
(general relativistic version of the energy-momentum balance equation), along with the equation for the
estimate of mass,
under the background of above mentioned respective metrics, given by 
\begin{eqnarray}
T^{\mu\nu}_{;\nu}=0,~~{\rm where}~~T^{\mu\nu}=(P+\rho+U)u^{\mu}u^{\nu}+Pg^{\mu\nu},~~{\rm and}~~
\left(\rho u^\mu\right)_{;\mu}=0,
\label{tmu}
\end{eqnarray}
for a perfect fluid,
where $P$, $\rho$ and $U$ are respectively the pressure, mass density and internal energy density of the matter
(as well as the magnetic field, if present) and $u^\mu$ is its 4-velocity. 

\section{Measuring spin of black holes from accretion properties}

Measuring spin of black holes, i.e. the Kerr parameter, of observed black hole sources
is a challenging job, while mass is comparatively easier to measure. The main methods for spin
measurements are: (1) fitting the thermal continuum from accretion disks, (2) inner disk reflection modeling, 
(3) modeling the QPOs. The first two methods are more popular, however often producing
contradictory results \cite{narayan,fabian}. The main reason for the third method not being 
as popular is the uncertainly behind the origin of QPOs. Nevertheless, I myself explored QPOs 
to determine the spin of stellar mass black holes and neutron stars \cite{m09,m13} by a unified scheme. 

Another approach, also related to the accretion properties, is to establish a relation 
between the mass and spin of black holes and, hence, measuring spin by supplying the mass \cite{prl}. 
Although the event horizon is a function of the mass and spin of black holes, 
it does not serve the purpose as it is not unique for all black holes. Hence,
in order to relate the mass and spin, one may plan to rely upon the properties of accretion disks.
Following Novikov and Thorne \cite{nt73}, the
solutions of equations (\ref{tmu}) for insignificant radial velocity
for the metric given by equation (\ref{kerrm}) with $Q=0$, 
the luminosity of the disk around a rotating black hole can be given by 
 \begin{equation}
L=\int_{r_{ISCO}}^{r_{out}}F~dr = \int_{r_{ISCO}}^{r_{out}}[7 \times 10^{26}~{\rm erg~cm^2~s^{-1}}](\dot{m} m^{-1})r_*^{-3}B ^{-1} C ^{-1/2}Q~dr,
\label{lum}
\end{equation}
where $m=M/M_{\odot}$, $\dot{m}=\dot{M}/\dot{M}_{Edd}$,
$\dot{M}$ and $\dot{M}_{Edd}$ respectively being the mass accretion
rate and the Eddington accretion rate, $r$ is the arbitrary distance
in the accretion disk from the black hole, $r_{out}$ and $r_{ISCO}$ are respectively the outer and 
inner radii of the disk, and $B, C, Q$ are functions of $M$ and $a$ (see \cite{nt73} for exact expressions).
Since $L$ is (approximately) fixed for a given class of black holes,
equation (\ref{lum}) reveals to be a 3-parameter algebraic equation, relating $M$, $a$ and $\dot{m}$.
Hence, if $\dot{m}$ is known, equation (\ref{lum}) is useful to measure $a$ for a known $M$.

Stellar mass black hole sources mainly exhibit two (extreme) classes of accretion flow:
(1) an optically thick and geometrically thin accretion disk (Keplerian flow) with 
$L\sim 10^{37}-10^{38}$ erg/sec and $\dot{m}\sim 0.1$, (2) an optically thin and 
geometrically thick accretion disk (sub-Keplerian flow) with 
$L\lesssim 10^{35}$ erg/sec and $\dot{m}\lesssim 10^{-4}$. On the other hand, 
supermassive black holes are classified into many groups, e.g., LINER, Seyfert, FR-I, FR-II, again
based on their respected luminosities and the class of, e.g., quasars harboring 
the respective black holes. 

Eighty quasars with ̇known respective $\dot{m}$, $M$ and $L$ are given by Ref. 39. 
Hence, using equation (\ref{lum}), I predict each of their $a$, some of which are listed in Table 1. 
It clearly shows that $a$ spans the range from
a very low to a high value, without clustering around a particular $a$. This proves that
there is {\it no bias} in this calculation.
Interestingly, our theory shows that a black hole may form with $a\rightarrow 1$ and, 
then, $a$ may exceed unity by accreting matter; furthermore, leading
to the formation of a naked singularity, which in turn may enlighten the issue of cosmic censorship.

\begin{table}[h]
\caption{Spins of supermassive black holes with known optical luminosity, $L_{\rm opt}$,
in units of erg/sec, accretion rate $\dot{\tilde{m}}$ in units of $M_\odot$/year and mass.}
\begin{center}
\begin{tabular}{|c|c|c|c|c|}

\hline
$\rm Object$ & $\rm log(m)$ & $\rm log(\dot{\tilde{m}})$ & $\rm log(L_{opt})$ & $a$ \\
\hline
$\rm 1425+267$ & $\rm 9.53$ & $\rm 0.07$ & $ \rm 45.55$ & $0.977$\\
\hline
$\rm 1048-090$ & $\rm 9.01$ & $\rm 0.30$ & $ \rm 45.45$ & $0.781$\\
\hline
$\rm 0947+396$ & $\rm 8.71$ & $\rm 0.19$ & $ \rm 45.20$ & $0.582$\\
\hline
$\rm 2251+113$ & $\rm 8.86$ & $\rm 0.66$ & $ \rm 45.60$ & $0.444$\\
\hline
$\rm 1226+023$ & $\rm 9.01$ & $\rm 1.18$ & $ \rm 46.03$ & $0.222$\\
\hline
$\rm 1302-102$ & $\rm 8.76$ & $\rm 0.92$ & $ \rm 45.71$ & $0.043$\\
\hline
$\rm 2112+059$ & $\rm 8.85$ & $\rm 1.16$ & $ \rm 45.92$ & $-0.057$\\
\hline
\end{tabular}
\end{center}
\end{table}

\section{Massive, magnetized, rotating white dwarfs in general relativity and modified 
general relativity}

\subsection{General Relativity}

Type~Ia supernovae (SNeIa) are believed to result from the
violent thermonuclear explosion of a carbon-oxygen white dwarf,
when its mass approaches the famous Chandrasekhar limit of
$1.44M_\odot$. For the discovery of the mass-limit of white dwarfs, 
S. Chandrasekhar was awarded the Nobel Prize in Physics in 1983 along with
W. A. Fowler who contributed towards the formation of the chemical elements in the universe.
SNIa is used as a standard candle in understanding
the expansion history of the universe \cite{perl99}. 
This very feature led to the Nobel Prize
in Physics in 2011, awarded to S. Perlmutter, B. P. Schmidt
and A. G. Riess, who, by observing distant SNeIa,
discovered that the universe is undergoing an accelerated expansion.

However, some of these SNeIa are highly over-luminous, e.g. SN 2003fg, SN 2006gz, SN 2007if, SN 2009dc 
\cite{howel,scalzo},
and some others are highly under-luminous, e.g. SN 1991bg, SN 1997cn, SN 1998de, SN 1999by
\cite{1991bg,taub2008}. 
The luminosity of the former group (super-SNeIa) implies highly super-Chandrasekhar 
white dwarfs, having mass $2.1-2.8M_\odot$, as their most plausible progenitors \cite{howel,scalzo}. 
While, the latter group (sub-SNeIa) predicts that the progenitor mass could be as low
as $ \sim M_\odot$ \cite{1991bg}.  
The models attempted to explain them so far entail caveats.

In a series of papers, with my collaborators, I argued that highly magnetized white dwarfs could be
as massive as inferred from the above observations \cite{prd,prll,jcap14}. As a
strong magnetic field corresponds to non-negligible magnetic pressure and 
magnetic density controlling the equilibrium structure of the star, apart
from its possible quantum mechanical effects (Landau quantization), a general
relativistic treatment is more useful to describe such white dwarfs \cite{jcap15a}. This is
more so as their radius could be less than $500$ km --- they are much more compact
than their nonmagnetic counterparts --- in particular for poloidally dominated magnetic field configurations.
Hence, the effects of GR are important to take into account to describe highly magnetized 
white dwarfs, just like in the case of neutron stars. The formalism to describe such a star, which 
may be highly spheroidal in shape, depending upon the field strength, has been 
elaborated in, e.g., Ref. 49. These authors have also made available a
code developed to describe highly magnetized neutron stars in GR, namely {\it XNS}, 
to the public. This is basically a solver-code of equations (\ref{tmu}) in hydro/magnetostatic conditions
for a given set of spacetime metric and EoS.

Furthermore, my collaborators and myself modified the {\it XNS} code in order to make it
appropriate for white dwarfs. I showed that poloidally dominated white dwarfs are 
smaller in size (with an equatorial radius substantially smaller than $1000$ km), whereas
toroidally dominated ones have larger radii \cite{jcap15a}. However, either of them could be 
significantly super-Chandrasekhar.

\begin{figure}
\includegraphics[width=0.9\linewidth]{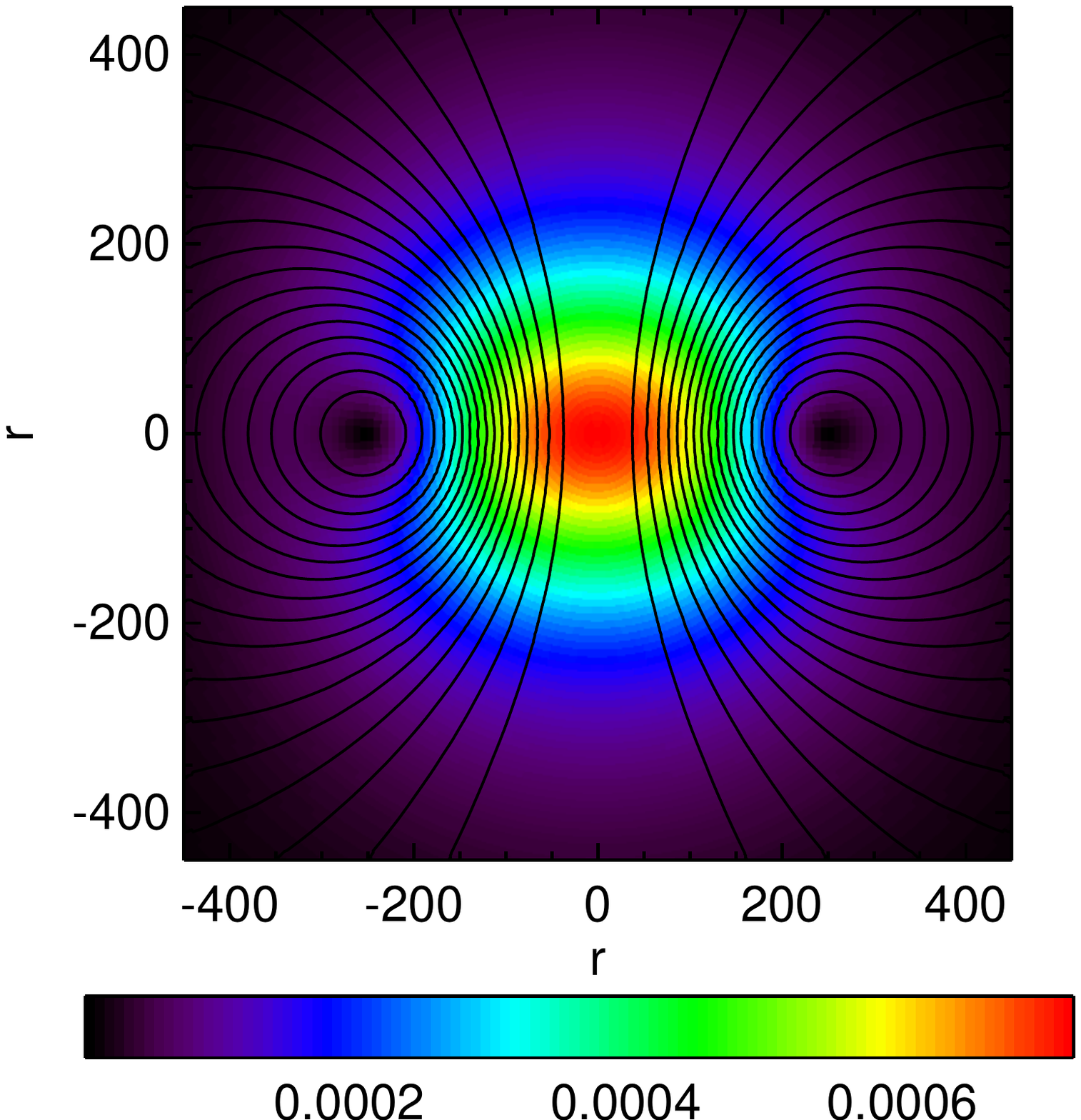}
\caption[]{Illustration of poloidal magnetic field geometries, when the black contours are the magnetic surfaces. 
The field is in units of $10^{14}$ G and $r$ is in units of 1.48km. 
}
\label{wd1}
\end{figure}

\begin{figure}
\includegraphics[width=0.9\linewidth]{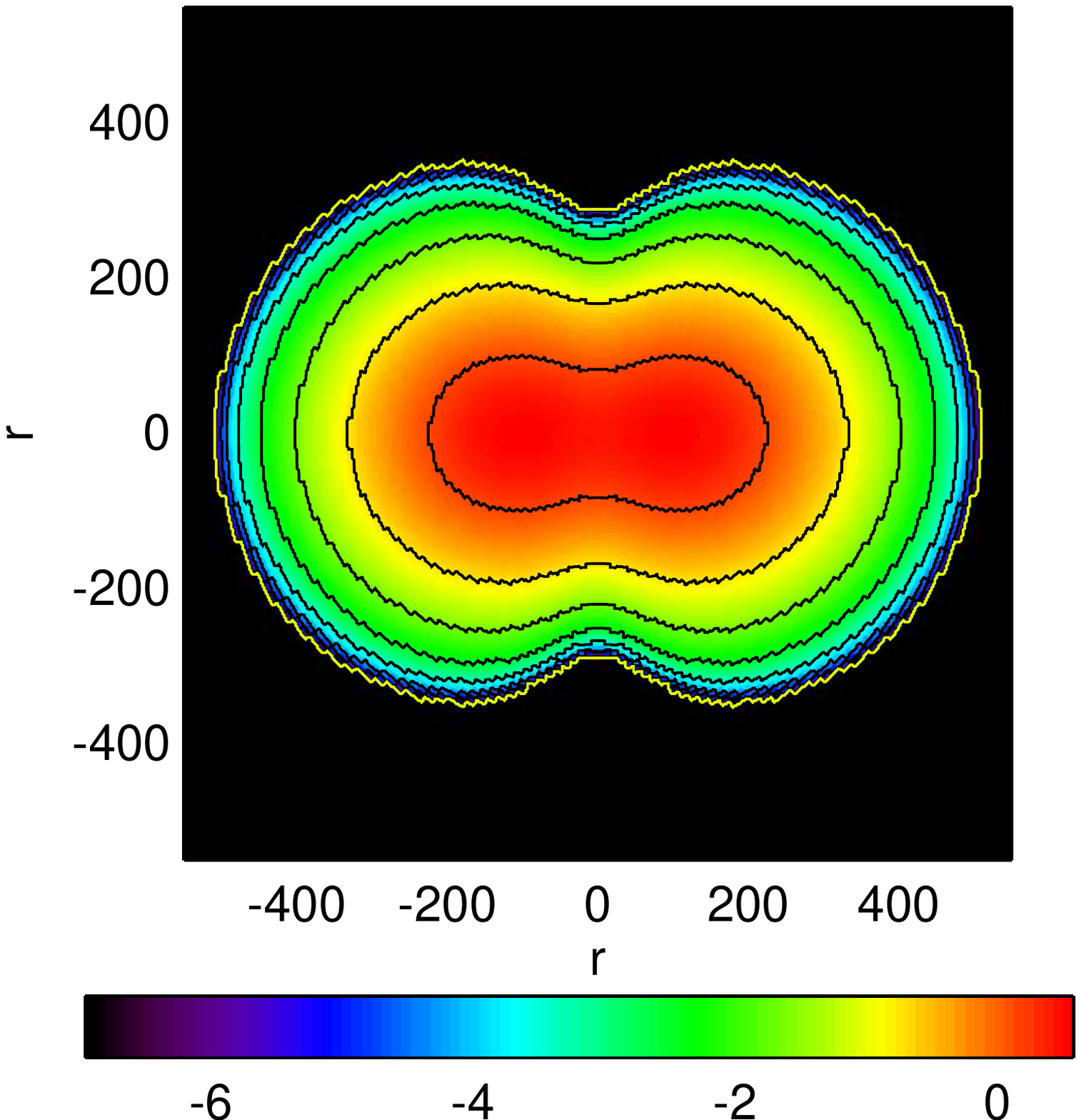}
\caption[]{The contour plots of density in logarithmic scale, when density is in units of $10^{10}$ gm/cc
for differentially rotating white dwarfs having
poloidal magnetic fields with $M = 1.95M_\odot$, equatorial radius $746$ km,  the ratio of polar to
equatorial radii $0.564$. 
}
 \label{wd2}
\end{figure}

Subsequently, along with my collaborator, I explored the effects of rotation in white dwarfs and 
found that rotation alone can increase the mass only upto $\sim 1.8M_\odot$ before
rotational instability may set in \cite{ostriker}, while the combined effects
of rotation and magnetic field lead to much more massive white dwarfs. Indeed, 
white dwarfs should be both rotating and magnetized in general.
Figure \ref{wd1} shows a typical geometry of poloidal magnetic field and Fig. \ref{wd2} shows the shape 
of a poloidally dominated white dwarf having mass $\sim 2M_\odot$ and radius $\sim 750$ km. Furthermore,
Table 1 shows that the mass of a differentially rotating, toroidally dominated white dwarf could exceed
$3M_\odot$. Note that I restrict the ratios of kinetic to gravitational energies (KE/GE) and 
magnetic to gravitational energies (ME/GE) to $0.2$ in order to assure stability. 
Higher values still could reveal equilibrium white dwarfs with much higher masses (see, \cite{jcap15a,sathya}).

\begin{table}
\caption{Differentially rotating configurations having purely toroidal magnetic
field, with changing maximum magnetic field $B_{max}$ within 
white dwarfs, when central angular velocity $\Omega_c=30.42$ sec$^{-1}$ fixed.
$r_e$ and $r_p$ are respectively the equatorial and polar radii.
}
\centering
\resizebox{\linewidth}{!}{
\begin{tabular}{|c|c|c|c|c|c|c|}
\hline
$B_{max} (10^{14}~{\rm G})$  & $M (M_\odot)$ & $r_e ({\rm km})$ & $\Omega_{eq} ({\rm sec^{-1}})$ & KE/GE & ME/GE & $r_p/r_e$\tabularnewline
\hline
\hline
0 & 1.769 & 1410 & 2.990 & 0.126 & 0 & 0.613 \tabularnewline
\hline
2.299 & 1.959 & 1676 & 2.180 & 0.132 & 0.046 & 0.603 \tabularnewline
\hline
2.996 & 2.318 & 2171 & 1.339 & 0.136 & 0.108 & 0.583 \tabularnewline
\hline
3.584 & 3.159 & 3322 & 0.593 & 0.132 & 0.203 & 0.584 \tabularnewline
\hline
\end{tabular}}
\end{table}

\subsection{Modified General Relativity}

Magnetized white dwarfs are unable to explain the under-luminous SNeIa mentioned above.
There are however some proposed models, with caveats, to describe them.
For example, numerical simulations of the merger of two sub-Chandrasekhar white dwarfs
reproduce the low power of under-luminous SNeIa, however the simulated light-curves fade
slower than that suggested by observations.

A major concern, however, is that a large array
of models is required to explain apparently the same phenomena, i.e., triggering of thermonuclear explosions
in white dwarfs. Why nature would seek mutually uncorrelated scenarios to exhibit sub- and super-SNeIa?
This is where the idea of modifying GR steps in into the context of white dwarfs, which unifies
the sub-classes of SNeIa by a single underlying theory. 

Let me consider, for the present purpose, the simplistic Starobinsky model 
\cite{staro} defined as $f(R)=R+\alpha R^2$, when $\alpha$ is a constant. 
However, similar effects could also be obtained in other, physically more sophisticated, theories, where
$\alpha$ (or effective-$\alpha$) is varying (e.g., with density). Now, on extremizing the action equation (\ref{action})
for Starobinsky's model, 
one obtains the modified field equation of the form
\begin{equation}
G_{\mu\nu}+\alpha X_{\mu\nu}= \frac{8\pi G T_{\mu\nu}}{c^4},
\label{modfld}
\end{equation}
where $T_{\mu\nu}$ contains only the matter field (non-magnetic star)
and $X_{\mu\nu}$ is a function of $g_{\mu\nu}$, $R_{\mu\nu}$ and $R$ (see \cite{jcap15b} for details).

Here, I seek perturbative solutions of equation (\ref{modfld}) (see, e.g., \cite{capo}), 
such that $\alpha R\ll 1$. Furthermore, I consider the hydrostatic equilibrium condition
so that $g_{\nu r}\nabla_\mu T^{\mu\nu}=0$, 
with zero velocity and $\nabla_\mu$ the covariant derivative. Hence, I
obtain the differential equations for mass $M_\alpha (r)$, pressure $P_\alpha (r)$
(or density $\rho_\alpha (r)$)
and gravitational potential $\phi_\alpha (r)$, of spherically symmetric white dwarfs
(which is basically the set of {\it modified} Tolman-Oppenheimer-Volkoff (TOV) equations). 
For $\alpha=0$, these  equations reduce to TOV equations in GR.

I supply EoS, obtained 
by Chandrasekhar \cite{chandra35}, as $P_0=K\rho_0^{1+(1/n)}$, where $P$ and $\rho$ of Ref. 53
are replaced by $P_0$ and $\rho_0$ respectively ($\alpha=0$: GR) in the spirit of perturbative approach. 
This form of EoS is valid for
extremely low and high densities, where $n$ is the polytropic index and $K$ a dimensional constant. 
The boundary conditions are: $M_\alpha (0)=0$ and $\rho_\alpha (0) =\rho_c$, where $\rho_c$ is the central 
density of the white dwarf. Note that a particular $\rho_c$ corresponds to a particular $M_\alpha$ and 
radius $R_\alpha$ 
of white dwarfs. Hence, by varying $\rho_c$ from $2\times 10^5$ gm/cc to $10^{11}$ gm/cc, I construct 
the mass-radius relation. 

Figures 1(a) and 1(b) show
that all three $M_\alpha-\rho_c$ curves for $\alpha<0$ overlap with the $\alpha=0$ curve in the low density region. 
However, with the increase 
of $\alpha$, the region of overlap recedes to a lower $\rho_c$. Modified GR
effects become important and visible at $\rho_c \gtrsim 10^8,~4\times 10^7$ and $2\times 10^6$ gm/cc, 
for $\alpha = 2\times 10^{13}~ {\rm cm^2}$, $8\times 10^{13}~ {\rm cm^2}$ and 
$10^{15}~{\rm cm^2}$ respectively. For a given $\alpha$, with the increase of $\rho_c$, $M_\alpha$ 
first increases, reaches a maximum and then decreases, like the $\alpha=0$ (GR) case. With the 
increase of $\alpha$, maximum mass $M_{\rm max}$ decreases and
for $\alpha=10^{15}~{\rm cm}^2$ it is highly sub-Chandrasekhar ($0.81M_\odot$).
This reveals that modified GR has a tremendous impact on white dwarfs.
In fact, $M_{\rm max}$ for all the chosen $\alpha>0$ is sub-Chandrasekhar, ranging $1.31-0.81M_\odot$. 
This is a remarkable finding since it establishes that even if $\rho_c$s for 
these sub-Chandrasekhar white dwarfs are lower than the conventional value at which SNeIa are usually 
triggered, an attempt 
to increase the mass beyond $M_{\rm max}$ with increasing $\rho_c$ will lead to a gravitational instability. 
This presumably will be followed by a 
runaway thermonuclear reaction, provided the core temperature increases sufficiently due to collapse. 
Occurrence of such thermonuclear runway reactions, triggered at densities as low as $10^6$ gm/cc,
has already been demonstrated \cite{runaway}. Thus, once $M_{\rm max}$ is approached, a SNIa is expected to trigger just 
like in the $\alpha=0$ case, explaining the sub-SNeIa \cite{1991bg,taub2008}, 
like SN 1991bg mentioned above. 

\begin{figure}[h]
\begin{center}
\includegraphics[angle=0,width=10cm]{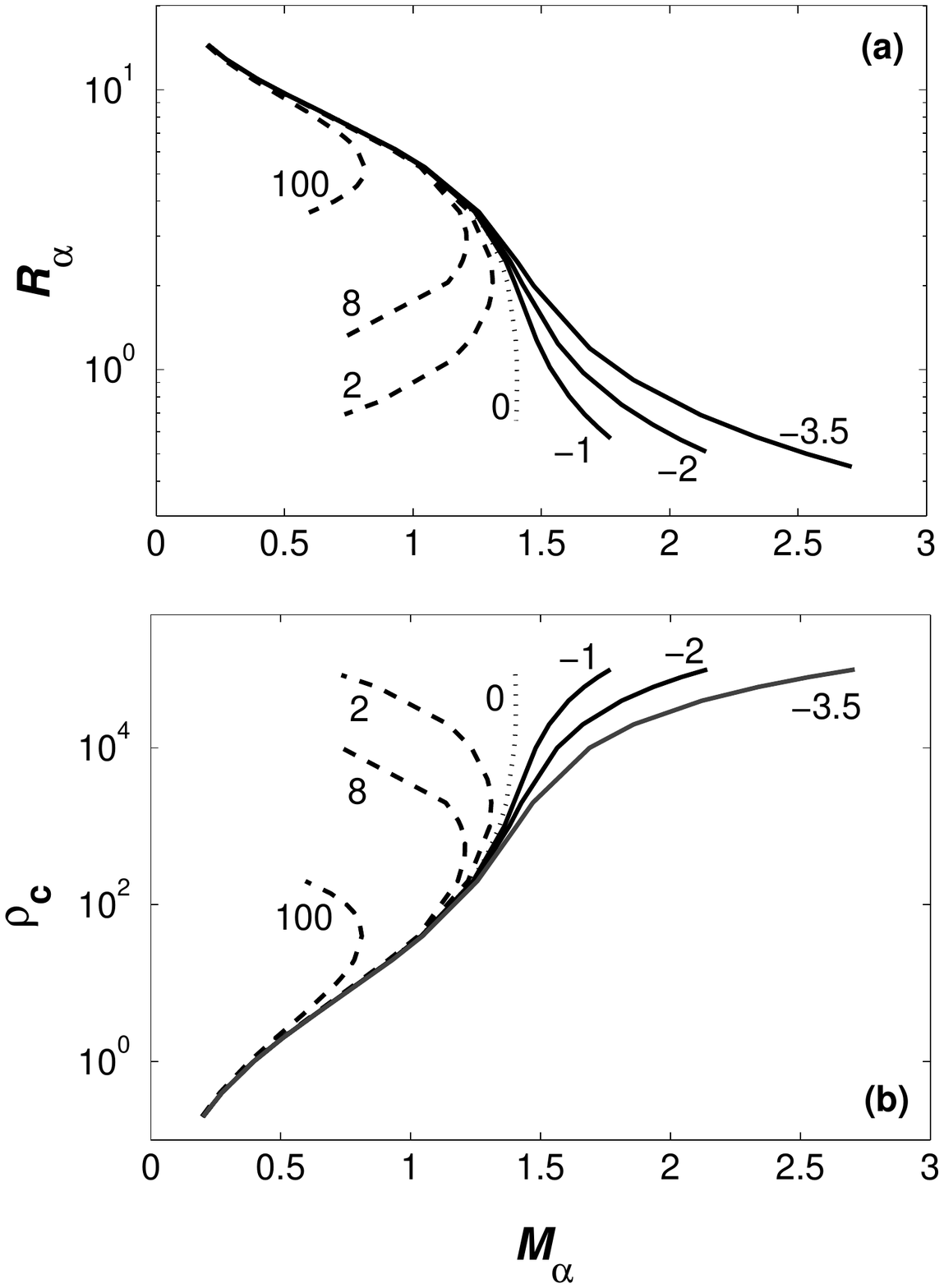}
\caption[]{Unification diagram for SNeIa:
(a) mass-radius relations, (b) variation of $\rho_c$ with $M_\alpha$.
The numbers adjacent to the various lines denote 
$\alpha/(10^{13}~{\rm cm^2})$. 
$\rho_c$, $M_\alpha$ and $R_\alpha$ are in units of $10^6$ gm/cc, $M_\odot$ and 1000 km respectively.
  }
\label{fig}
\end{center}
\end{figure}

For $\alpha<0$ cases, Fig. 1(b) shows that for $\rho_c>10^8$ gm/cc, the 
$M_\alpha-\rho_c$ curves deviate from the GR curve due to modified GR effects. 
Note that $M_{\rm max}$ for all the three cases corresponds to $\rho_c=10^{11}$ gm/cc, 
an upper-limit chosen to avoid possible neutronization. 
Interestingly, all values of $M_{\rm max}$ are highly super-Chandrasekhar, ranging from 
$1.8-2.7M_\odot$. 
Thus, while the GR effect is very small, modified GR effect could
lead to $\sim 100\%$ increase in the limiting mass of white dwarfs. 
The corresponding values of $\rho_c$ are large enough to initiate thermonuclear reactions, e.g.
they are larger than $\rho_c$ corresponding to $M_{\rm max}$ of $\alpha=0$ case, whereas the respective 
core temperatures are expected to be similar. This 
explains the entire range of the observed super-SNeIa mentioned above \cite{howel,scalzo},
assuming the furthermore gain of mass above $M_{\rm max}$ leads to SNeIa.


Tables 2 and 3 ensure the perturbative validity of the solutions. Recall that 
I solve the modified TOV equations only up to ${\cal O} (\alpha)$. Since the product 
$\alpha R$ is first order in $\alpha$, I replace $R$ in it by the zero-th order Ricci scalar 
$R^{(0)} = 8\pi(\rho^{(0)} - 3P^{(0)})$, which is Ricci scalar obtained in GR ($\alpha=0$). 
For the perturbative validity of the entire solution, 
$|\alpha R^{(0)}|_{\rm max} \ll 1$ should hold 
true. Next, I consider $g_{tt}^{(0)}/g_{tt}$ and $g_{rr}^{(0)}/g_{rr}$ (ratios of $g_{\mu\nu}$-s 
in GR and those in modified GR up to ${\cal O} (\alpha)$),
which should be close to unity for the validity of perturbative method \cite{orelana}. Hence,
$|1-g_{tt}^{(0)}/g_{tt}|_{\rm max} \ll 1$ and 
$|1-g_{rr}^{(0)}/g_{rr}|_{\rm max} \ll 1$ should both hold true.
Tables 2 and 3 indeed show that all the three measures quantifying perturbative validity are at least $2-3$ 
orders of magnitude smaller than 1.

\begin{table}[]
\caption{Measure of validity of perturbative solutions for $\alpha>0$ corresponding to $M_{\rm max}$ in Fig. \ref{fig}.}
\begin{center}
\small
\begin{tabular}{|c|c|c|c|}

\hline 
$\alpha/(10^{13}~{\rm cm}^2)$  & $|\alpha R^{(0)}|_{\rm max}$ & $|1-g^{(0)}_{tt}/g_{tt}|_{\rm max}$ & $|1-g^{(0)}_{rr}/g_{rr}|_{\rm max}$ \\ 
\hline

2 & $7.4\times 10^{-5}$ &  $6.8\times 10^{-5}$ & $2.0\times 10^{-4}$ \\ 

8 & $7.4\times 10^{-5}$ & $6.8\times 10^{-5}$ & $2.0\times 10^{-4}$ \\

100 & $7.4\times 10^{-5}$ & $6.9\times 10^{-5}$ & $2.0\times 10^{-4}$  \\ \hline

\end{tabular}

\end{center}

\end{table}

\begin{table}[]
\caption{Measure of validity of perturbative solutions for $\alpha<0$ corresponding to $M_{\rm max}$ in Fig. \ref{fig}.}
\begin{center}
\small
\begin{tabular}{|c|c|c|c|c|c|c|c|}

\hline 
$\alpha/(10^{13}~{\rm cm}^2)$  & 
$|\alpha R^{(0)}|_{\rm max}$ & $|1-g^{(0)}_{tt}/g_{tt}|_{\rm max}$ & $|1-g^{(0)}_{rr}/g_{rr}|_{\rm max}$ \\ 
\hline

-1 & 0.00184 & 0.0016 & 0.0052 \\ 

-2 &  0.00369 & 0.0031 & 0.0108 \\



-3.5 & 0.00646 & 0.0052 & 0.0199 \\ \hline

\end{tabular}

\end{center}

\end{table}

\subsubsection{Possible effect of density dependent model parameter leading to chameleon-like theory}
I now justify that the effects of modified GR based on a more sophisticated calculation, invoking
an (effective) $\alpha$ that varies explicitly with density (and effectively becomes negative), are likely to converge to those 
described above with constant $\alpha$.
Note that even though $\alpha$ is assumed to be constant within
individual white dwarfs here, there is indeed an implicit
dependence of $\alpha$ on $\rho_c$, particularly of the liming mass white dwarfs presumably leading to
SNeIa, as is evident from Fig. \ref{fig}(b). This indicates the existence
of an underlying chameleon effect. This trend is expected to emerge self-consistently in a varying-$\alpha$ theory.

Let me consider a possible situation where $\alpha$ varies explicitly with density and 
try to relate it with the above results. Note that the super-SNeIa occur mostly in young stellar 
populations consisting of massive stars (see, e.g., \cite{howel}),
while the sub-SNeIa occur in old stellar
populations consisting of low mass stars (see, e.g., \cite{gonz}). The massive stars with
higher densities are likely to collapse to give rise to
super-Chandrasekhar white dwarfs, which would subsequently
explode to produce super-SNeIa. The low mass
stars with lower densities would be expected to collapse
to give rise to sub-Chandrasekhar white dwarfs, which would
probably end in sub-SNeIa. Now, let me assume
a functional dependence of $\alpha$ on density such that there are two terms ---
one dominates at higher densities, while the other dominates at lower densities. Hence,
when a massive, high density star collapses, it yields results similar to our $\alpha < 0$ cases;
while when a low mass, low density star collapses, it leads to results
like our $\alpha > 0$ cases. Thus, the same functional
form of $\alpha$ could lead to both super- and sub-Chandrasekhar
limiting mass white dwarfs, respectively. Note that the final
mass of the white dwarf would depend on several factors,
such as, $\rho_c$ and the density gradient in
the parent star, etc. Interestingly, this description invoking a variation of
$\alpha$ with density is essentially equivalent to invoking
a so-called ``chameleon-$f(R)$ theory", which can pass solar
system tests of gravity (see, e.g., \cite{cham2}).
This is so because $\alpha$ is a function of density, which in turn
is a function of $R$ and, hence, introducing a
density (and hence $R$) dependence into $\alpha$ is equivalent to choosing an
appropriate (more complicated) $f(R)$ model of gravity.
Therefore, even when one invokes a more self-consistent variation of $\alpha$
with density, it does not invalidate the results of the
constant-$\alpha$ cases, rather is expected to complement the 
picture.

On a related note, I would like to mention that the order of magnitude of $\alpha$ is different
between that in typical white dwarfs ($\alpha \sim 10^{13}$ ${\rm cm^2}$, as used above) and in neutron stars
($\alpha \sim 10^9$ ${\rm cm^2}$, e.g. \cite{eksi,capo}). This again basically stems from
the fact that there is an underlying chameleon effect which causes
$\alpha$ to be different in different density regimes. Note that neutron stars are
much denser than white dwarfs and, hence, have a higher value
of curvature $R$. Now, the quantity $\alpha R$ would have a
similar value in both neutron stars and white dwarfs in the
perturbative regime. Hence, due to their higher curvature, neutron stars
will harbor a smaller value of $\alpha$ compared to white dwarfs. Roughly, neutron
stars are $10^4$ times denser than white dwarfs and, hence,
$\alpha_{neutron-star}$ is $10^4$ times smaller than $\alpha_{white-dwarf}$.


\section{Summary}

In the last several decades, relativistic astrophysics has turned out to be a highly important
branch in astrophysics. In this branch, many major astrophysical discoveries are still taking place in 
the contexts of black holes, quasars, neutron stars, white dwarfs, X-ray binaries, gamma-ray bursts, 
particle acceleration, the cosmic background, dark matter, dark energy etc., even 100 years after
Einstein's discovery of GR, which is the basic building block for them. The present article
has touched upon some of the underlying latest astrophysical problems and their possible 
resolutions. It has been revealed that while Einstein's gravity itself is indispensable to uncover
modern high energy astrophysical problems, modified Einstein's gravity also appears to be playing an important
role behind certain phenomena and, in general, to explain astrophysical processes.\\ \\

\noindent {\large\bf Acknowledgment}\\ \\
\noindent I am thankful to Indrani Banerjee, Mukul Bhattacharya, Upasana Das, Chanda J. Jog, Subroto Mukerjee, A. R. Rao,
Prateek Sharma and Sathyawageeswar Subramanian, for continuous discussions on the topics covered in this article.


\begin{thebibliography}{10}

\bibitem{ein} A. Einstein, SPAW, 844 (1915).

\bibitem{sch} K. Schwarzschild, AbhKP, 189 (1916).

\bibitem{kerr} R. Kerr, Phys. Rev. Lett. \textbf{11,} 237 (1963).

\bibitem{bl} R. H. Boyer, \& R. W. Lindquist, JMP \textbf{8,} 265 (1967).

\bibitem{hr} J. B. Hartle, \& K. S. Thorne, ApJ \textbf{153,} 807 (1968).

\bibitem{cook} G. B. Cook, S. L. Shapiro, \& S. A. Teukolsky, ApJ \textbf{424,} 823 (1994).

\bibitem{xns} N. Bucciantini, \& L. Del Zanna, A\&A \textbf{528,} A101 (2011).

\bibitem{hulse} R. A. Hulse, \& J. H. Taylor, ApJ \textbf{195,} 51 (1975).

\bibitem{staro} A. A. Starobinsky, Phys. Lett. B \textbf{91,} 99 (1980).

\bibitem{capo} A. V. Astashenok, S. Capozziello and S. D. Odintsov, JCAP \textbf{12,} 040 (2013).

\bibitem{eksi} S. Arapo\u{g}lu, C. Deliduman, \& K. Y. Ek\c{s}i, JCAP \textbf{7,} 020 (2011).

\bibitem{nat} P. B. Demorest, {\it et al.}, Nature \textbf{467,} 1081 (2010).

\bibitem{sc} J. Antoniadis, {\it et al.}, Science \textbf{340,} 448 (2013).

\bibitem{weissenborn} S. Weissenborn, D. Chatterjee, \& J. Schaffner-Bielich, 
Phys. Rev. C \textbf{85,} 065802 (2012).

\bibitem{whittenbuary} D. L. Whittenbury, J. D. Carroll, A. W. Thomas, K. Tsushima, \& J. R. Stone,
Phys. Rev. C. \textbf{89,} 065801 (2014).

\bibitem{pili} A. G. Pili, N. Bucciantini, \& L. Del Zanna, MNRAS \textbf{439,} 3541 (2014).

\bibitem{cheon} M.-K. Cheoun, C. Deliduman, C. Gungor, V. Keles, C. Y. Ryu, T. Kajino,
\& G. J. Mathews, JCAP \textbf{10,} 021 (2013).

\bibitem{bondi} H. Bondi, MNRAS \textbf{112,} 195 (1952).

\bibitem{michael} F. C.	Michel, Ap\&SS \textbf{15,} 153 (1972).

\bibitem{nt73} I. D. Novikov, \& K. S. Thorne, in Black Holes, ed. C. DeWitt \& B. S.
DeWitt (New York: Gordon and Breach), 343 (1973).

\bibitem{ss73} N. I. Shakura, R. A. Sunyaev, A\&A \textbf{24,} 366 (1973).

\bibitem{sel76} S. L. Shapiro, A. P. Lightman, \& D. M. Eardley, ApJ \textbf{204,} 187 (1976).

\bibitem{ny94}
R. Narayan, I. Yi, ApJ \textbf{428,} L13 (1994).

\bibitem{pw80} B. Paczy\'nsky, \& P. J. Wiita, A\&A \textbf{88,} 23 (1980).

\bibitem{m02} B. Mukhopadhyay, ApJ \textbf{581,} 427 (2002).

\bibitem{liang80} E. P. T. Liang, \& K. A. Thompson, ApJ \textbf{240,} 271 (1980).

\bibitem{c90} S. K. Chakrabarti, ApJ \textbf{350,} 275 (1990).

\bibitem{gp98} C. F. Gammie, \& R. Popham, ApJ \textbf{498,} 313 (1998).

\bibitem{belo07} W.-X. Chen, \& A. M. Beloborodov, ApJ \textbf{657,} 383 (2007).

\bibitem{sasha11} A. Tchekhovskoy, R. Narayan, \& J. C. McKinney, MNRAS \textbf{418,} 79 (2011).

\bibitem{jon12} J. C. McKinney, A. Tchekhovskoy, \& R. D. Blandford, MNRAS \textbf{423,} 3083 (2012).

\bibitem{jon14} J. C. McKinney, A. Tchekhovskoy, A. Sadowski, \& R. Narayan, MNRAS \textbf{441,} 3177 (2014).

\bibitem{livrel} A. de Felice, \& S. Tsujikawa, Liv. Rev. Rel. \textbf{13,} 3 (2010).

\bibitem{narayan} J. E.	McClintock, at al., CQG \textbf{28,} 114009 (2011).

\bibitem{fabian} R. C. Reis, A. C. Fabian, R. R. Ross, \& J. M. Miller, MNRAS \textbf{395,} 1257 (2009).

\bibitem{m09} B. Mukhopadhyay, ApJ \textbf{694,} 387 (2009).

\bibitem{m13} B. Mukhopadhyay, D. Bhattacharya, \& P. Sreekumar, IJMPD \textbf{21,} 1250086 (2012).

\bibitem{prl} I. Banerjee, \& B. Mukhopadhyay, Phys. Rev. Lett. \textbf{111,} 061101 (2013).

\bibitem{davis-laor} S. W. Davis, \& A. Laor, ApJ \textbf{728,} 98 (2011).







\bibitem{perl99}
S. Perlmutter, {\it et al.}, Astrophys. J. \textbf{517,} 565 (1999).

\bibitem{howel}
D. A. Howell, {\it et al.}, Nature \textbf{443,} 308 (2006).


\bibitem{scalzo}
R. A. Scalzo, {\it et al.}, Astrophys. J. \textbf{713,} 1073 (2010).


\bibitem{1991bg}
A. V. Filippenko, {\it et al.}, Astron. J. \textbf{104,} 1543  (1992).

\bibitem{taub2008}
S. Taubenberger, {\it et al.}, Mon. Not. R. Astron. Soc. \textbf{385,} 75 (2008).

\bibitem{prd} U. Das, \& B. Mukhopadhyay, Phys. Rev. D \textbf{86,} 042001 (2012).

\bibitem{prll} U. Das, \& B. Mukhopadhyay, Phys. Rev. Lett. \textbf{110,} 071102 (2013).

\bibitem{jcap14} U. Das, \& B. Mukhopadhyay, JCAP \textbf{06,} 050 (2014). 

\bibitem{jcap15a} U. Das, \& B. Mukhopadhyay, JCAP \textbf{05,} 016 (2015). 

\bibitem{pilli}	A. G. Pili, N. Bucciantini, \& L. Del Zanna, MNRAS \textbf{439,} 3541 (2014). 

\bibitem{ostriker} J. P. Ostriker, \& F. D. A. Hartwick, ApJ \textbf{153,} 797 (1968). 

\bibitem{sathya} S. Subramanian, \& B. Mukhopadhyay, MNRAS \textbf{454,} 752 (2015). 


\bibitem{jcap15b} U. Das, \& B. Mukhopadhyay, JCAP \textbf{05,} 045 (2015). 

\bibitem{chandra35} S. Chandrasekhar, MNRAS \textbf{95,} 207 (1935).

\bibitem{runaway}
I. R. Seitenzahl, C. A. Meakin, D. M. Townsley, D. Q. Lamb, \& J. W. Truran, ApJ \textbf{696,} 515 (2009).

\bibitem{orelana}
M. Orellana, F. Garc\'{i}a, F. A. Teppa Pannia, \& G. E. Romero, Gen. Rel. Grav. \textbf{45,} 771 (2013).

\bibitem{gonz}
S. Gonz\'{a}lez-Gait\'{a}n, {\it et al.}, ApJ \textbf{727,} 107 (2011).

\bibitem{cham2} T. Faulkner, M. Tegmark, E. F. Bunn, Y. Mao, Phys. Rev. D \textbf{76,} 063505 (2007).




\end{thebibliography}
\end{document}